\documentclass[useAMS,usenatbib]{mn2e}

\usepackage{amsmath}
\usepackage{amssymb}
\usepackage[dvipsnames]{xcolor}
\usepackage{graphicx}
\usepackage{graphics}
\usepackage{hyperref}
\hypersetup{
     colorlinks   = true,
     citecolor    = blue
}

\def\apj{{ ApJ}}
\def\apjl{{ApJL}}

\def\aap{{ A\&A}}

\def\mnras{{ MNRAS}}

\def\nat {{ Nature}}

\def\pra{{ Phys. Rev. A}}
\def\pre{{ Phys. Rev. E}}

\def\prl{{ Phys. Rev. Lett}}

\def\mr{\mathrm}

\def\mc{\mathcal}
\def\d{\mr{d}}
\def\b{\mathbf}

\def\omp{\omega_{\rm p}}
\def\hz{\hat{\mathbf{z}}}

\newcommand{\lara}[1]{{\langle#1\rangle}}

\title[FRB propagation in local environment]{Imprint of local environment on fast radio burst observations}
\author[Lu \& Phinney]{Wenbin
  Lu$^{1}$\thanks{wenbinlu@caltech.edu} and E. Sterl Phinney$^1$
\\ $^{1}$Theoretical Astrophysics, and Walter Burke Institute for Theoretical Physics, Mail Code
350-17, Caltech, Pasadena, CA 91125, USA 
}

\begin{document}
\label{firstpage}
\maketitle

\begin{abstract}
When fast radio burst (FRB) waves propagate through the local ($\lesssim \rm1\,pc$) environment of the FRB source, electrons in the plasma undergo large-amplitude oscillations. The finite-amplitude effects cause the effective plasma frequency and cyclotron frequency to be dependent on the wave strength. The dispersion measure and rotation measure should therefore vary slightly from burst to burst for a repeating source, depending on the luminosity and frequency of the individual burst. Furthermore, free-free absorption of strong waves is suppressed due to the accelerated electrons' reduced energy exchange in Coulomb collisions. This allows bright low-frequency bursts to propagate through an environment that would be optically thick to low-amplitude waves. Given a large sample of bursts from a repeating source, it would be possible to use the deficit of low-frequency and low-luminosity bursts to infer the emission measure of the local intervening plasma and its distance from the source. Information about the local environment will shed light on the nature of FRB sources. 
\end{abstract}


\section{Introduction}
Many bright fast radio bursts (FRBs) with isotropic luminosities of order $10^{45} \rm\,erg\,s^{-1}$ or higher were found by recent observations, e.g., FRB 010724 \citep{2007Sci...318..777L, 2019MNRAS.482.1966R}, FRB 160102 \citep{2018MNRAS.475.1427B}, 180817.J1533+42 \citep{2019Natur.566..230C}, 181016 \citep{2019MNRAS.488.2989F}, 181112 \citep{2019Sci...366..231P}, 190523 \citep{2019Natur.572..352R}. Statistical studies show that the FRB luminosity function possibly extends up to $\sim 10^{47}\rm\,erg\,s^{-1}$ \citep{2019ApJ...883...40L}. For coherent linearly polarized waves, the amplitude of the electric field at a distance $r$ is related to the isotropic luminosity $L$ by
\begin{equation}
  \label{eq:12}
  E_0 = \left(\frac{2L}{r^2 c}\right)^{1/2},
\end{equation}
The dimensionless wave strength is defined as \citep{1971ApJ...165..523G, 2014ApJ...785L..26L}
\begin{equation}
  \label{eq:11}
  a = \frac{E_0 e}{m \omega c} = 7.2\times10^{-3} L_{45}^{1/2}
  r_{17}^{-1} \nu_9^{-1},
\end{equation}
where $L = 10^{45} L_{45}\rm\,erg\,s^{-1}$ (the fiducial value corresponds to a flux of $18\rm\, Jy\, GHz$ at redshift 1), $r = 10^{17}r_{17}\rm\,cm$, $\nu = 10^9\nu_9\,$Hz is the wave frequency, and $c$ is the speed of light. The field amplitude and wave strength are smaller by a factor of $\sqrt{2}$ for a circularly polarized wave of the same luminosity. In this paper, we explore some observable signatures of large-amplitude wave propagation. We use Gaussian CGS units and the widely adopted subscript notation of $X_n \equiv X/10^n$.

\section{Propagation of Finite-Amplitude Waves}
Finite amplitude wave-plasma interactions have been intensively studied \citep[e.g.,][and references therein]{Akhiezer56, 1970PhFl...13..472K, 1971PhRvL..27.1342M, 1972PhRvL..29.1731M,  1975PhLA...55..227F, 1976A&A....47...97D, 1981PhRvA..24.2773C, 2006RvMP...78..309M} in the context of pulsars and high-power pulsed lasers. These studies all restricted themselves to steady-state periodic solutions. \citet{1971PhFl...14..999N, 1971ApJ...165..523G} investigated the acceleration of individual particles by a pulse, but did not fully consider the effect of plasma. While protons stay nearly at rest, electrons' motion in the longitudinal direction generates a kinematically important electrostatic wake field. Here we revisit the calculation, emphasizing certain points relevant for the propagation of an intense wave packet of finite length in an electron-ion plasma, for both circular and linear polarizations. While finishing this paper, we learned of the work of \citet{yangzhang20}, who investigated the propagation of finite-amplitude waves (for circular polarization only) and also concluded that the dispersion measure of FRBs from the local environment depends on the wave strength.

We start from the Maxwell equations
\begin{equation}
    \nabla \cdot \b{E} = 4\pi e (n-n_0),\ \  \nabla \cdot \b{B} = 0,
\end{equation}
\begin{equation}
    \nabla \times \b{E} = -c^{-1}\partial_t \b{B},\ \ \nabla \times \b{B} =c^{-1}( \partial_t \b{E} + 4\pi e n \b{v}),
\end{equation}
where $n$ is the electron density, $n_0$ is the constant proton density, $e$ is electron charge, and $\b{v}$ is electron speed. A divergence-free B-field can be written in terms of a vector potential $\b{A}$ and the E-field can be described by $\b{A}$ and the scalar potential $V$,
\begin{equation}
    \b{B} = \nabla \times \b{A},\ \ \b{E} = -c^{-1}\partial_t\b{A} - \nabla V.
\end{equation}
Adopting Coulomb gauge $\nabla\cdot \b{A}=0$, we write Gauss's law in the form of a Poisson equation
\begin{equation}\label{eq:Poisson}
    \nabla^2V = -4\pi e(n-n_0),
\end{equation}
and Ampere's law in the form of a wave equation
\begin{equation}
    (c^{-1}\partial_t^2 - c\nabla^2)\b{A} + \partial_t\nabla V = 4\pi e n \b{v}. 
\end{equation}
In the following, we take the system to be one-dimensional with all variations only in the $z$ direction, and consider a transverse plane wave propagating along the $z$ axis, e.g., for the linearly polarized case, $\b{E}=\b{E_0}\exp[i(kz-\omega t)]$. For such a 1-d system, the Coulomb gauge condition implies that $\b{A}$ is purely transverse, and longitudinal terms arise only from the gradient of the scalar potential $V$. We thus identify the $\partial_t\nabla V$ term as the longitudinal (along $\hz$) current
\begin{equation}
\label{eq:longit}
    \partial_t\nabla V = 4\pi e n v_z \hz.
\end{equation}
This can be seen by taking the divergence of the Ampere's law, which gives
\begin{equation}
    \nabla\cdot (4\pi e n \b{v} - \partial_t \nabla V) = 0.
\end{equation}
Thus, the wave equation can be decoupled into a longitudinal component (eq. \ref{eq:longit}) and transverse components
\begin{equation}\label{eq:transv}
    (c^{-1}\partial_t^2 - c\nabla^2)\b{A} = 4\pi e n \b{v}_\perp. 
\end{equation}
It follows that the oscillating part of the vector potential $\b{A}$ is purely transverse. Note that there could be a constant magnetic field in the plasma, but we ignore it for the moment since the cyclotron frequency is much smaller than the wave frequency considered here. 

To solve the potentials $\b{A}$ and $V$, we need the electron's equation of motion, which is described by the Lagrangian
\begin{equation}
    \mc{L} = -mc^2/\gamma + e(\b{A}\cdot \b{v}/c - V),
\end{equation}
where $\gamma \equiv (1-v^2/c^2)^{-1/2}$ is the Lorentz factor. Since the Lagrangian does not explicitly depend on $x$ and $y$ (the coordinates in the transverse direction), these two components of the Canonical momentum $\b{p} = m\b{u} + e\b{A}/c$ ($\b{u}\equiv \gamma\b{v}$ is the four-velocity) are conserved. We consider the case where the amplitude of the electromagnetic wave slowly ramps up from zero to maximum on a timescale much longer than the wavelength. We ignore initial thermal motion of the particles and hence set $p_x = p_y = 0$, which means
\begin{equation}
\label{eq:transverse_u}
    u_x = -\frac{e}{mc} A_x,\ \ u_y = -\frac{e}{mc}A_y.
\end{equation}
The $z$ component of the vector potential is $A_z=0$ in the absence of a static magnetic field. Thus, the equation governing the vector potential (eq. \ref{eq:transv}) becomes 
\begin{equation}\label{eq:wave_eq}
\begin{split}
    (c^2\partial_z^2 - \partial_t^2)\b{A} &= -(4\pi e n_0 - \partial_z^2V) \frac{-e \b{A}}{\gamma m} \\
        &= \frac{\omp^2}{\gamma} \left(1 - \frac{\partial_z^2V}{4\pi en_0}\right) \b{A},
\end{split}
\end{equation}
where we have used the Poisson equation to eliminate $n$ and defined the plasma frequency $\omp = \sqrt{4\pi e^2n_0/m}$. The scalar potential equation (\ref{eq:longit}) can be rewritten as
\begin{equation}
\label{eq:V_evolution}
    \partial_t\partial_{z} V = 4\pi e n_0 \left(1 - \frac{\partial_z^2V}{4\pi en_0}\right) v_z.
\end{equation}
This can be understood that in the limit $\partial_z^2V/(4\pi e n_0)\ll 1$ (a strong wake field hasn't developed yet), the longitudinal electric field $\partial_zV$ grows with time due to charge separation as a result of longitudinal electron motion $v_z$ (since protons are not moving).

We restrict ourselves to the non-relativistic case of $a=eA_0/(mc^2)\ll 1$, $A_0$ being the amplitude of the vector potential. In this limit, $v_z$ is of order $\mc{O}(a^2)$ and the Lorentz factor is approximately given by $u_x$ and $u_y$, i.e.
\begin{equation}
    \gamma \approx (1 + u_x^2/c^2 + u_y^2/c^2)^{1/2} = (1 + e^2A^2/m^2c^4)^{1/2}.
\end{equation}
The detailed evolution of the scalar potential is complex, but eq. (\ref{eq:V_evolution}) shows that $\partial_z^2V/4\pi e n_0$ is at most of order $v_z$. Therefore, as a reasonable approximation, we ignore the $(1-\partial_z^2V/4\pi e n_0)$ factor\footnote{Including the $(1-\partial_z^2V/4\pi e n_0)$ factor will introduce a numerical factor in front of the $e^2A^2/2m^2$ term, but our eq. (\ref{eq:approx_wave_eq}) captures the non-linear effects qualitatively.} in the wave equation (\ref{eq:wave_eq}) and obtain
\begin{equation}
\label{eq:approx_wave_eq}
    (c^2\partial_z^2 - \partial_t^2)\b{A} \approx \omp^2 \left(1 - \frac{e^2A^2}{2m^2c^4}\right)\b{A}.
\end{equation}
A rough, intuitive way of understanding the above result is that electrons are accelerated by the wave electric field to an averaged Lorentz factor of $\lara{\gamma}$ and hence the plasma frequency is reduced by a factor of $\lara{\gamma}^{-1/2}$ due to the relativistic correction to the inertial mass of the electron.

\subsection{Implications}
The simplest case is for a circularly polarized wave of amplitude $|A|=A_0$. Using the dimensionless wave strength $a= eA_0/(mc^2)$, the dispersion relation for this case is
\begin{equation}
    \omega^2 - c^2k^2 = \omp^2 (1 - a^2/2), \ \ \mbox{for\ cir. pol.,}
\end{equation}
in agreement with \citet{Akhiezer56}. For linearly polarized polarized wave, since the nonlinear term involves $\cos^3\psi=(3\cos\psi + \cos3\psi)/4$, we adopt the ansatz $\b{A} = \b{A}_0(\cos\psi + C\cos3\psi)$, where $\psi = kz-\omega t$ is the phase and $|C|\ll 1$ is a constant. Then, the LHS of the wave equation is
\begin{equation}
    (c^2\partial_z^2 -\partial_t^2)\b{A} = (\omega^2-c^2k^2)\b{A}_0(\cos\psi + 9C\cos3\psi).
\end{equation}
If we only retain the lowest order terms $\mc{O}(a^2)$ and $\mc{O}(C)$, the RHS becomes
\begin{equation}
   \omp^2 \b{A}_0 [(1-3a^2/8)\cos\psi + (C-a^2/8)\cos3\psi].
\end{equation}
In order for the wave equation to hold for arbitrary phase $\psi$, the coefficients of the $\cos\psi$ and $\cos3\psi$ terms must be equal, so we obtain $C = -a^2/64$,
and hence the dispersion relation is
\begin{equation}
    \omega^2 - k^2 = \omp^2(1 - 3a^2/8),\ \ \mbox{for lin. pol.}
\end{equation}
The wave develops some small amplitude $\mc{O}(a^2)$ harmonics at frequency $3\omega$ in the transverse direction. The longitudinal oscillation is given by $u_z \approx (e^2A^2/2m^2c^3) \cos^2\psi$, which oscillates at frequency $2\omega$. Thus, it is a quasi-transverse wave mode (no purely transverse mode is possible for linear polarization). On the other hand, for multi-frequency propagation, e.g. two modes with $\psi_i=k_i z-\omega_i t$ (for $i=1, 2$), other than the linear terms $\cos\psi_i$, there will be non-linear $\mc{O}(a^2)$ terms related to $\cos^3\psi_1$, $\cos^2\psi_1 \cos\psi_2$, $\cos\psi_1\cos^2\psi_2$ and $\cos^3\psi_2$, thus involving frequencies $3\omega_1$, $2\omega_1 \pm \omega_2$, $2\omega_2\pm \omega_1$, and $3\omega_2$. The nonlinear evolution of wave packets with a continuous Fourier spectrum will be studied in a future work.

Here, we take linear polarization as the fiducial case (circular polarization gives similar results) and replace the wave strength $a$ by the isotropic luminosity $L$ and radius $r$ according to eq. (\ref{eq:11}), and then the dispersion relation can be written as
\begin{equation}
  \label{eq:28}
  \omega^2 = k^2 + \omp^2 (1 - \delta),\  \delta = 1.9\times10^{-5} L_{45} r_{17}^{-2} \nu_9^{-2}.
\end{equation}
The effect of a large amplitude wave is to reduce the dispersion measure (DM) of the local environment (at distance $r$) by a factor of $1-\delta$. The DM variation caused by the wave amplitude effects is given by
\begin{equation}
    \Delta \mr{DM} = \int\d r\, n(r) \delta(r) \simeq (6.3\times10^{-3} {\mr{pc\,cm^{-3}}})\, n_4 L_{45} r_{17}^{-1} \nu_9^{-2}.
\end{equation}
It can be shown that the same physics applies to magnetized plasma as well, in the sense that both the square of the plasma frequency and cyclotron frequency $\omega_B=eB/mc$ are reduced by a factor of $\lara{\gamma}^{-1}$, and hence the local rotation measure contribution $\propto \omega_p^2\omega_B$  is reduced by a factor of $1-2\delta$. These finite-amplitude effects may be noticeable for bright, low-frequency bursts with sufficient time and frequency resolution.

\section{Free-free absorption}
In the previous section, we have considered electron oscillations driven by finite-amplitude waves, ignoring the existence of protons (or ions). Occasionally, the electron may undergo a Coulomb collision with a proton and hence gain or lose energy from the electromagnetic field. Since on average energy-absorbing encounters dominate energy-emitting ones, this process is known as free-free (ff) absorption. It has been extensively discussed in the literature both classically and quantum mechanically \citep[e.g.,][and references therein]{1979rpa..book.....R, 1973PhRvA...8..804K, 1979PhRvA..20.1934S, 1994PhPl....1.4043D, 1995PhRvE..51.4778P, 2001PhRvE..64b6414B}.

For large wave amplitudes such that electrons' oscillatory speed $eA_0/mc$ is much larger than the thermal speed $\sqrt{k_{\rm B}T/m}$ at plasma temperature $T$ ($k_{\rm B}$ being the Boltzmann constant), the electron motion corresponds to a higher effective temperature $T_{\rm eff}\sim e^2A_0^2/(k_{\rm B}mc^2)\gg T$, so the ff absorption coefficient is suppressed by a factor of order $(T_{\rm eff}/T)^{-3/2}$. In this section, we discuss how the suppression of ff absorption affects the detection of low-frequency FRBs \citep[e.g.,][]{2019ApJ...874...72R}. We define a dimensionless wave amplitude $\xi_0 \equiv eA_0/\sqrt{mc^2k_{\rm B}T}$ (roughly the ratio between the oscillatory and thermal speeds), and then the ff absorption optical depth is given by
\begin{equation}
  \label{eq:41}
  \begin{split}
  \tau_{\rm ff}(\nu) &= \frac{8 \lara{\mr{ln}\,\Lambda}_{\xi_0}}{ 3(2\pi)^{1/2}} \frac{\mr{EM}\,e^6}{
  c(mk_{\rm B}T)^{3/2}\nu^2} F(\xi_0)\\
&= 3.0\times10^{16} \frac{\lara{\mr{ln}\,\Lambda}_{\xi_0} F(\xi_0)}{(T/\mr{K})^{3/2}
  (\nu/\mr{Hz})^2} \frac{\mr{EM}}{
  \mr{pc\,cm^{-6}}},
  \end{split}
\end{equation}
where $\mr{EM}= \int n^2 \d \ell$ is the emission measure, the average Coulomb logarithm is given by
\begin{equation}
  \label{eq:58}
\lara{\mr{ln}{\,\Lambda}}_{\xi_0} \approx
  \begin{cases}
\mr{ln} [3k_{\rm B}T(1 +
\xi_0^2)/h\omega],\mbox{ cir. pol.},\\
\mr{ln} [3k_{\rm B}T(1 + \xi_0^2/2)/h\omega], \mbox{ lin. pol.},
  \end{cases}
\end{equation}
and we find that the suppression factor $F$ can be fitted by the following simple forms (with maximum fractional error of 15\%)
\begin{equation}
  \label{eq:55}
  F(\xi_0) = 
  \begin{cases}
    (1 - 0.33\xi_0 + 0.83\xi_0^2)^{-3/2},\mbox{ cir. pol.},\\
    (1 + 0.15\xi_0 + 0.35\xi_0^2)^{-2.85/2},\mbox{ lin. pol.}
  \end{cases}
\end{equation}
A classical calculation of the suppression factor $F$ is provided later on in \S\ref{ff_suppression} (and our Figures are based on accurate numerical results). We note that, when the Coulomb logarithm is included, in the limit $\xi_0\gg 1$, the absorption coefficient roughly scales as $\alpha_{\rm ff}(\xi_0)\propto \xi_0^{-2.9}$ for circular polarization and $\propto \xi_0^{-2.75}$ for linear polarization. In the following, we first discuss the observational consequences of the modified absorption coefficient.


\subsection{Escape of low-frequency FRB radiation}\label{FRB_escaping}

\begin{figure*}
  \centering
\includegraphics[width = 0.47 \textwidth]{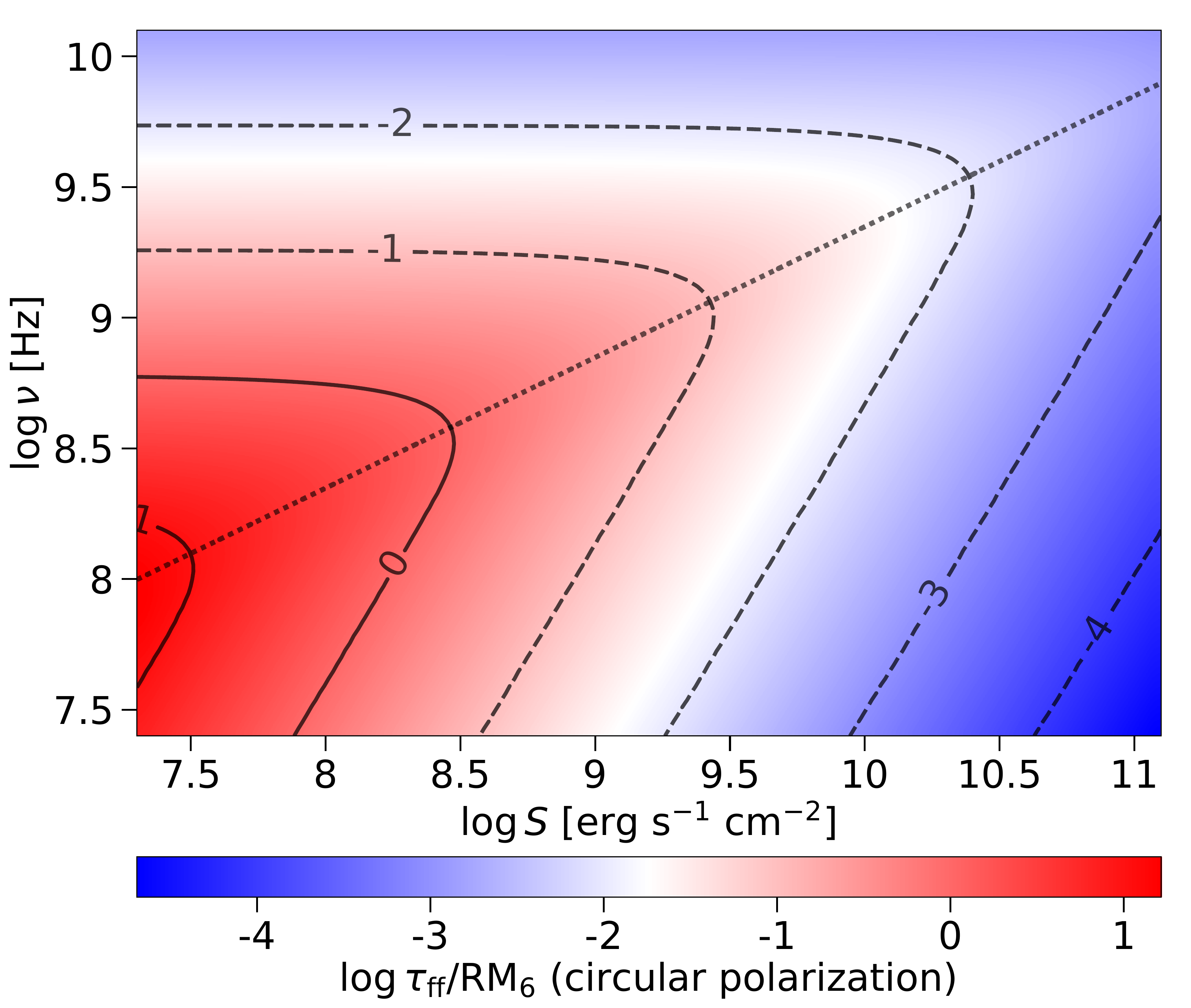}
\includegraphics[width = 0.47 \textwidth]{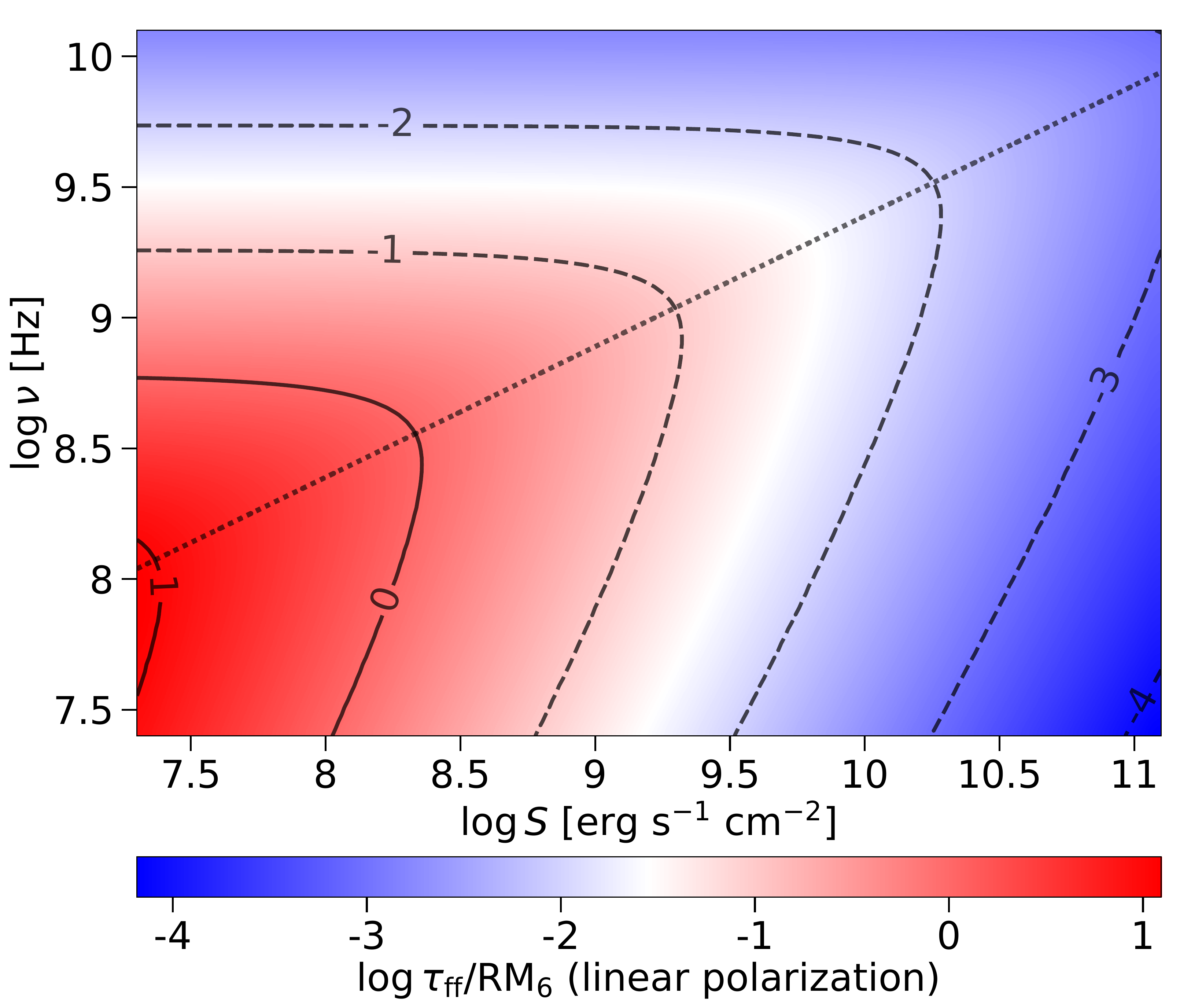}
\caption{The ff optical depth divided by $\mr{EM}_6\equiv \mr{EM}/10^6 \rm\,pc\,cm^{-6}$, as a function of wave energy flux $S = L/4\pi r^2$ and wave frequency $\nu$, for a plasma of temperature $T=10^4\rm\,K$. The \textit{left panel} is for circular polarization and the \textit{right panel} is for linear polarization. A reference energy flux is $S\approx 8\times10^9 \rm\, erg\, s^{-1}\,cm^{-2}$ for $L = 10^{45}\rm\,erg\,s^{-1}$ and $r = 10^{17}\rm\,cm$. The cyan dashed line shows the critical frequency at which the dimensionless strength $\xi_0 = 1.4$ (left panel) and $1.8$ (right panel), roughly corresponding to when the finite-amplitude effects become important.
}\label{fig:tauff}
\end{figure*}

In Fig. \ref{fig:tauff}, we show the ff optical depth divided by $\mr{EM}_6\equiv \mr{EM}/10^6 \rm\,pc\,cm^{-6}$ as a function of the wave energy flux $S = \lara{A^2}\omega^2/4\pi c=L/4\pi r^2$ and wave frequency $\nu$, for a plasma of temperature $T = 10^4\rm\,K$. We find that the ff optical depth scales as $\d\, \mr{log}\,\tau_{\rm ff}/\d\, \mr{log}\, \nu \simeq -2.1$ at $\nu\gg \nu_{\rm c}$ (when finite-amplitude effects are not important) and $\d\, \mr{log}\,\tau_{\rm ff}/\d \,\mr{log}\, \nu \simeq 0.9$ (cir. pol.) or $\simeq 0.75$ (lin. pol.) at $\nu\ll \nu_{\rm c}$, where the critical frequency $\nu_{\rm c}$ is roughly given by (for both polarizations)
\begin{equation}
  \label{eq:59}
  \nu_{\rm c} \simeq (2.3\mr{\,GHz})\, S_{10}^{1/2} T_4^{-1/2}.
\end{equation}
The implications are: (1) FRBs at frequency $\nu\lesssim \nu_{\rm c}$ are more likely to escape due to finite-amplitude suppression of ff absorption; (2) By monitoring a repeating source at different frequencies, it may be possible to find luminosity-dependent ff absorption frequency (where $\tau_{\rm ff} = 1$), which provides a way of probing the local environment of FRB sources.

\begin{figure}
  \centering
\includegraphics[width = 0.45 \textwidth]{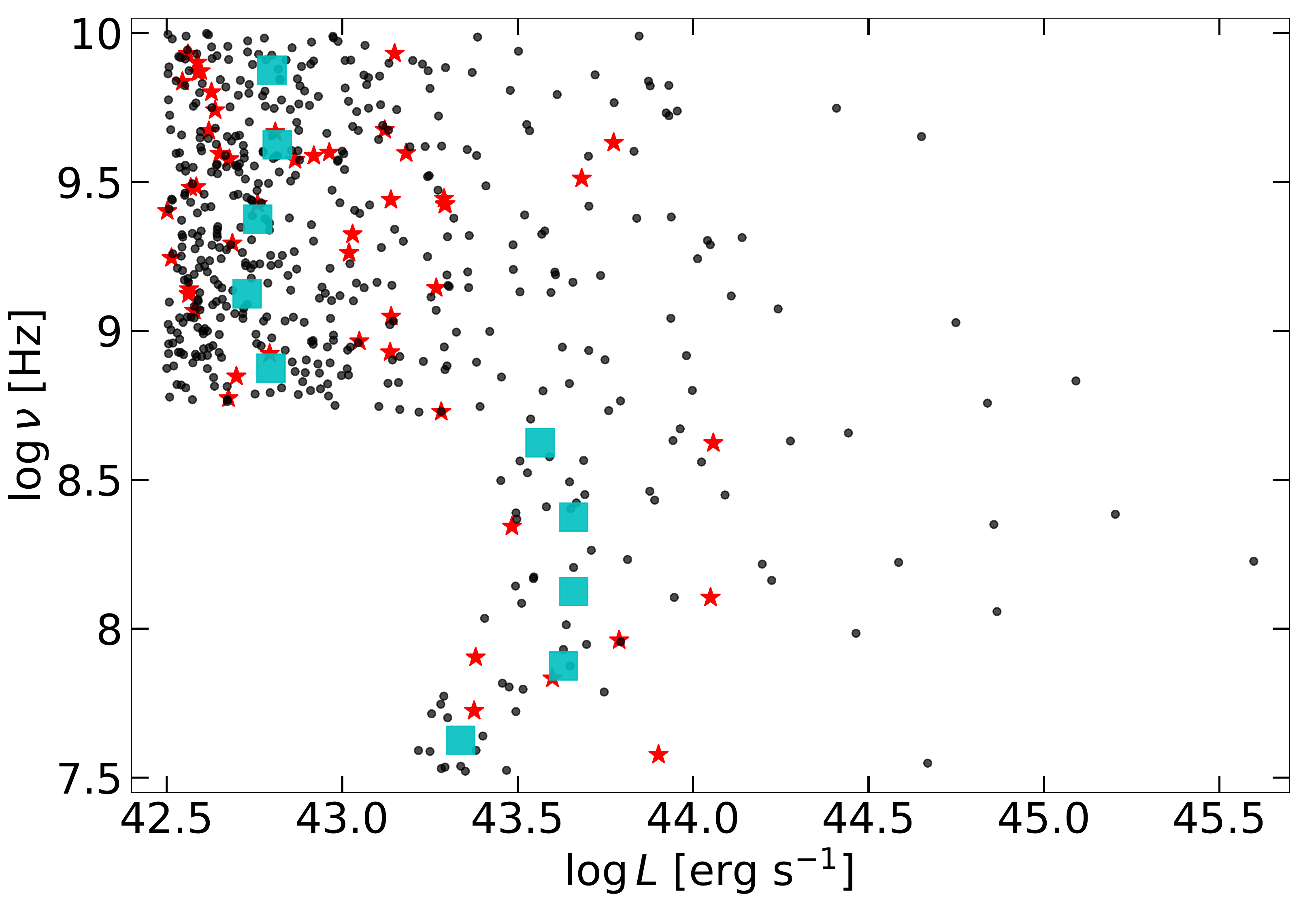}
\caption{Two mock samples of bursts from a repeating source collected at a wide range of frequencies from $30\,$MHz to $10\,$GHz. The larger sample (black dots) has size $N=500$ and the smaller sample (red stars) has $N=50$. The samples are generated by the following (highly simplified) conditions: the occurrence probability density is flat in log-frequency and power-law in luminosity, i.e. $\d P/\d \mr{\,log\,\nu} =\rm const.$, $\d P/\d \mr{\,log\,} L\propto L^{-\beta}$, and $\beta=1$. A burst is ``detected'' only when the ff optical depth $\tau_{\rm ff}<1$ for linear polarization (the results for circular polarization are similar). For the $N=500$ sample, we bin the ``detected'' bursts in log-frequency and show the median luminosity for each bin in cyan squares.
}\label{fig:scatter}
\end{figure}

To illustrate how this might be done, in Fig. \ref{fig:scatter}, we show two mock samples of individual bursts collected from a single repeating source. The repetition statistics of the best-studied FRB 121102 are still under debate \citep{2017ApJ...850...76L, 2019ApJ...877L..19G, 2019ApJ...882..108W}. So to demonstrate the qualitative results, we consider the following simplified conditions: the occurrence probability density is flat in log-frequency and a power-law in luminosity, i.e. $\d P/\d \mr{\,log\,\nu} =\rm const.$, $\d P/\d \mr{\,log\,} L\propto L^{-\beta}$, and take $\beta=1$ (consistent with current observational constraints). The minimum/maximum frequencies are $\mr{log}\,\nu_{\rm min} \mr{\,[Hz]} = 7.5$ and $\mr{log}\,\nu_{\rm max} \mr{\,[Hz]} = 10$. We note that broadband spectra are not needed for individual bursts, but our method does require the emission mechanism for each repeater to be operating at a sufficiently wide range of frequencies. The minimum/maximum luminosities are $\mr{log}\,L_{\rm min}\,\mr{[erg\,s^{-1}]} = 42.5$ and $\mr{log}\,L_{\rm max} \,\mr{[erg\,s^{-1}]} = 46$. Although bursts with much lower luminosities than $L_{\rm min}$ have been detected from FRB 121102, our results do not depend on the choice of $L_{\rm min}$.  All bursts are assumed to be linearly polarized. We consider a burst as ``detected'' only when the ff optical depth $\tau_{\rm ff}<1$.

For the $N=500$ sample, we bin the ``detected'' bursts in log-frequency and show the median luminosity for each bin in cyan squares. The boundary at $\tau_{\rm ff}=1$ should be smoother in reality, but the median luminosity in each bin is not strongly affected. Notice the jump in median luminosity at $\nu_{\rm jump}\simeq 500\rm\, MHz$, which is the frequency at which low-amplitude ($\xi_0\rightarrow 0$) bursts are ff absorbed. This can be used to infer the $\rm EM$ of the absorbing plasma
\begin{equation}
  \label{eq:60}
  \mr{EM} \simeq 8\times10^5\frac{\mr{pc}}{\mr{cm}^{4}} \frac{10}{
    \lara{\mr{ln}\,\Lambda}} T_4^{1.5}
  \left(\frac{\nu_{\rm jump}}{500\mr{\,MHz}}\right)^2,
\end{equation}
which is close to the true value used in the simulation: $10^6\rm\,pc\,cm^{-4}$ (for $T = 10^4\rm\,K$). The median luminosity $L_{\rm jump}\simeq 10^{43.5}\rm\, erg\,s^{-1}$ in the bin right below $\nu_{\rm jump}$ can be plugged into eq. (\ref{eq:59}) to infer the distance between the plasma and the FRB source 
\begin{equation}
  \label{eq:56}
  r \simeq (7\times10^{16}\mr{\,cm})\, T_4^{-1/2} L_{\rm
    jump,43.5}^{1/2} \left(\frac{\nu_{\rm jump}}{500\mr{\,MHz}}
    \right)^{-1},
\end{equation}
which is again close to the actual value used in the simulation: $10^{17}\rm\,cm$. Qualitatively similar, but less accurate, conclusions can be drawn from the $N=50$ sample. We also note that the above choices of $\mr{EM}$ and $r$ are only for illustrative purpose, whereas in reality different FRB sources may be located in extremely diverse environments. Finally, we note that the degeneracy in the inferred parameters ($\rm EM$ and $\rm r$) with the plasma temperature $T$ cannot be eliminated, unless we have additional information, such as the H$\alpha$ intensity.

\subsection{Free-free absorption for finite-amplitude waves}\label{ff_suppression}
In this subsection, we provide a classical calculation of the suppression factor $F$ (defined in eq. \ref{eq:41}) for ff absorption.

We are mainly concerned with Coulomb collisions with impact parameter $b\ll b_{\rm max} \equiv v/\omega$, which means the scattering time is much less than a wave period and hence the electromagnetic field does not change during the interaction\footnote{For large impact parameter $b\gtrsim b_{\rm max} \equiv v/\omega$, the Coulomb scattering occurs over multiple wave periods and the electron adiabatically adjusts to the slowly evolving combined electromagnetic$+$Coulomb fields and hence there is little energy exchange. }. This is the impulse approximation. We also adopt a minimum\footnote{Another possible $b_{\rm min}$ is $b_{90}=e^2/mv^2$ at which the electron is deflected by $90^{\rm o}$ \citep{1979rpa..book.....R}. For electron speeds larger than the thermal speed at $T\sim 10^4\rm\,K$, the quantum lower limit is more stringent. } impact parameter $b_{\rm min} = h/mv$ ($h$ being the Planck constant), because an electron with known momentum $mv$ cannot be described classically if closer to a proton than $b_{\rm min}$. The precise values of $b_{\rm max}$ and $b_{\rm min}$ do not strongly affect our results since they enter the final absorption coefficient through the Coulomb logarithm $\mr{ln}(b_{\rm max}/b_{\rm min})$.

To keep the expressions concise, we take $c=k_{\rm B}=1$ in this subsection, which can be easily recovered based on dimensional consideration. In the non-relativistic case (ignoring $\mc{O}(a^2)$ terms), the electron has canonical momentum
\begin{equation}
  \label{eq:34}
  \b{p} = \b{q} + e\b{A},
\end{equation}
where $\b{q}=m\b{v}$ is the kinetic momentum and $\b{A}(\psi)$ is the vector potential of the electromagnetic wave (such that the electric field is $\b{E} = -\omega \b{A}'$). In the absence of Coulomb collisions, the canonical momentum is conserved to the order $\mc{O}(a^2)$. Under an additional Coulomb potential $V$, the Hamiltonian is given by
\begin{equation}
  \label{eq:29}
  H = \frac{(\b{q}-e\b{A})^2}{2m} + eV = \frac{q^2}{2m} + eV +
  \frac{e^2A^2}{2m} - \frac{\b{q}\cdot \b{A}}{m}.
\end{equation}
We assume that protons remain motionless and hence the Coulomb scattering is elastic\footnote{For weak scatterings which dominate the ff absorption, the electron momentum kick is $\sim e^2/bv=(b_{90}/b) m v < mv$, the proton will receive a kick velocity $\sim (m v/m_p)(b_{90}/b)$ and hence energy of $(m/m_p) (b_{90}/b)^2 mv^2/2 \ll mv^2/2$.}. Thus, $q^2/2m + eV$ is conserved during the collision, $e^2A^2/2m$ does not change under the impulse approximation, and the energy gain/loss per scattering is given by
\begin{equation}
  \label{eq:25}
  \Delta E(\b{q}_1\rightarrow \b{q}_2) = -\frac{e}{m}\b{Q}\cdot \b{A},\ \ \b{Q}
  \equiv \b{q}_2-\b{q_1} = \b{p}_2 - \b{p}_1.
\end{equation}
The Rutherford formula for differential cross-section is
\begin{equation}
  \label{eq:30}
  \frac{\d \sigma}{\d \Omega} = \frac{e^4}{m^2 v^4 (1 -
    \cos\Theta)^2},
\end{equation}
where $\Theta$ is the deflection angle given by $\cos\Theta = \b{q}_1\cdot \b{q}_2/q^2$ (where $q = q_1=q_2$). Using $Q=|\b{q}_2-\b{q_1}|$ as defined in eq. (\ref{eq:25}), it can be recast in a simpler form 
\begin{equation}
  \label{eq:31}
  \frac{\d \sigma}{ \d \Omega} = 4m^2 e^4/Q^4.
\end{equation}

In the following, we consider the unperturbed electron distribution function to be Maxwellian in canonical momentum (taking Boltzmann constant $k_{\rm B}=1$)
\begin{equation}
  \label{eq:32}
  f(\b{p}) = (2\pi mT)^{-3/2} \mr{e}^{-p^2/2mT},
\end{equation}
which is normalized such that $\int f \d^3p = 1$. The rate of transition from state $\b{p}_1$ to the phase-space volume $\d^3p_2$ near $\b{p}_2$ is denoted as $R(\b{p}_1, \b{p}_2) \d^3 p_2$. At a given wave phase $\psi$, there is a one-to-one map between canonical momentum $\b{p}$ and kinetic momentum $\b{q}$ (eq. \ref{eq:34}), so we have $\d^3p = d^3 q$. Then $R$ is related to the differential cross-section (eq. \ref{eq:31}) as follows
\begin{equation}
  \label{eq:33}
  R(\b{p}_1, \b{p}_2) \d^3 q_2 = \delta(q_1-q_2) \d q_2 \cdot n v_1
  \frac{\d \sigma }{\d \Omega} \d \Omega_{q_2},
\end{equation}
where $\delta(q_2-q_1)$ is the Dirac delta-function as a result of elasticity. Making use of $\d^3 q_2 = q_2^2 \d q_2 \d \Omega_{q_2}$ and eq. (\ref{eq:31}) for $\d \sigma/\d \Omega$, we obtain the differential scattering rate
\begin{equation}
  \label{eq:35}
  R(\b{p}_1, \b{p}_2) = \delta(q_2 - q_1)  \frac{4nme^4}{ q Q^4}. 
\end{equation}
One can easily see that $R(\b{p}_1, \b{p}_2)$ is symmetric between $\b{p}_1$ and $\b{p}_2$ as a result of the symmetry of Coulomb scattering.

Therefore, the volumetric heating rate (in $\rm erg\,s^{-1}\,cm^{-3}$) due to ff absorption is given by
\begin{equation}
  \label{eq:36}
  \dot{U} = \frac{1}{2} n \int \d^3 p_1 \int \d^3 p_2 [f(\b{p}_1) - f(\b{p}_2)] \Delta E
  \cdot R(\b{p}_1, \b{p}_2),
\end{equation}
where $f(\b{p}_1)-f(\b{p}_2)$ accounts for the cancellation between forward $\d^3p_1\rightarrow \d^3p_2$ and backward $\d^3p_2\rightarrow \d^3p_1$ processes, and the factor of $1/2$ is due to double-counting in the full integral. In the weak scattering limit, we have $\Delta E\ll T$ and hence $f(\b{p}_1) - f(\b{p}_2) \propto 1 - \mr{e}^{-\Delta E/T} = \Delta E/T$. Then the integral can be carried out by using $\d^3 p_2 = d^3 q_2 = q_2^2 \d q_2 \d \Omega_{q_2}$, and we obtain
\begin{equation}
  \label{eq:37}
  \begin{split}
  \dot{U} =\, & \frac{2n^2 e^6 }{ (2\pi)^{3/2} (mT)^{5/2}} \int \d^3 p_1
  \mr{exp}\left(\frac{-p_1^2}{2mT}\right) \\
&\int \d q_2 \delta (q_2-q_1) q_2
  \int \d \Omega_{q_2} \frac{(\b{Q}\cdot \b{A})^2}{Q^4}.
  \end{split}
\end{equation}
At a given moment, let $\b{A} = A\hz$ and we define $\cos \theta_q \equiv \b{A}\cdot \b{q}_1/(Aq_1)$. In the weak scattering limit, the deflection vector $\b{Q}$ lies in a plane perpendicular to $\b{q}_1$ and the amplitude is $Q = q_1 \sin \theta$. We write $\d \Omega_{q_2} = \sin \theta \d \theta \d\phi$, where $\theta$ is the angle between $\b{q}_1$ and $\b{q}_2$, and $\phi$ is the azimuthal angle (starting from the intersecting line between the Q-plane and the $\b{q}_1$--$\b{A}$ plane). Then, we obtain
\begin{equation}
  \label{eq:38}
  \b{Q}\cdot \b{A} = QA\sin \theta_q \cos\phi,
\end{equation}
and the $\d \Omega_{q_2}$ integral becomes
\begin{equation}
  \label{eq:39}
  \int \d \Omega_{q_2} \frac{(\b{Q}\cdot \b{A})^2}{Q^4} = \frac{A^2
    \sin^2\theta_q}{q_1^2} \int\d \phi
  \cos^2\phi \int  \frac{\d \theta}{\sin \theta}.
\end{equation}
The deflection angle $\theta$ is related to
the impact parameter $b$ by $\mr{tan}(\theta/2) = e^2/(bmv^2)\approx \theta/2$ (in the weak scattering limit), so the $\theta$ integral can be written as $\int \d \theta/ \sin\theta\approx \int_{b_{\rm min}}^{b_{\rm max}}\d b/b = \mr{ln}(b_{\rm max}/b_{\rm min})\equiv \mr{ln}\,\Lambda$, which is known as the Coulomb logarithm. The $\phi$ integral is trivial $\int_0^{2\pi}\d \phi \cos^2\phi = \pi$. Further more, we make use of $\b{A} = A\hz$, $\sin^2\theta_q = (p_1^2 - p_{1,z}^2)/q_1^2$, and $q_1^2 = (\b{p}_1 - e\b{A})^2$, and then the volumetric ff heating rate becomes
\begin{equation}
  \label{eq:42}
  \dot{U} = \frac{\mr{ln}\,\Lambda\, A^2 n^2 e^6}{(2\pi)^{1/2}
    (mT)^{5/2}} \int \d^3 p_1 
 \frac{(p_1^2 - p_{1,z}^2) \mr{e}^{-p_1^2/ 2mT}}{(p_1^2 - 2ep_{1,z}A +
   e^2 A^2)^{3/2}}.
\end{equation}
The integral can be carried out by using $\d^3p_1 = 2\pi p_1^2\d p_1\sin \theta_p\d \theta_p$, where $\theta_p$ is the angle between $\b{p}_1$ and $\b{A}$ (i.e. $p_{1,z} = p_1\cos\theta_p$). Note that, for simplicity, we have ignored the dependence of the Coulomb logarithm $\mr{ln}\,\Lambda = \mr{ln}(mv^2/h\omega)$ on the particle's momentum. In the small amplitude limit $eA\ll p_1$, the integral is elementary and the heating rate is
\begin{equation}
  \label{eq:43}
  \dot{U} = \frac{4}{3} \frac{(2\pi)^{1/2} \mr{ln}\,\Lambda\, A^2 n^2 e^6
    }{(mT)^{3/2}}.
\end{equation}
Averaging over the wave phase $\lara{A^2} = \int_0^\pi (\d\psi/\pi)A^2$, we obtain the Poynting flux $S = \lara{A^2}\omega^2/4\pi$. Then the ff absorption coefficient is given by 
\begin{equation}
  \label{eq:44}
  \alpha_{\rm ff} = \frac{\lara{\dot{U}}}{S} = \frac{8\, \mr{ln\,}\Lambda
    }{3(2\pi)^{1/2}} \frac{n^2 e^6}{(mT)^{3/2}\nu^2},
\end{equation}
which is independent of polarization and agrees with the result from the thermal ff emissivity combined with the Kirkhoff theorem \citep[see eq. 5.19a of][]{1979rpa..book.....R}.

For the case of finite-amplitude waves, the integral $I(\xi) \equiv (4\pi mT)^{-1}\int \d^3 p_1 (\cdots)$ in eq. (\ref{eq:42}) can be written as
\begin{equation}
  \label{eq:51}
  \begin{split}
  &I(\xi) = \frac{1}{(2\xi)^{3/2}}
\int_0^{\infty} \d x\, x^{5/2} \mr{e}^{-x^2} \int_{-1}^1 \frac{(1-y^2)\d y}{
  (t - y)^{3/2}},\\
&x = \frac{p_1}{\sqrt{2mT}},\ y = \cos \theta_p,\ \xi = \frac{eA}{
  \sqrt{2mT}},\ t = \frac{x^2 + \xi^2}{2x\xi}.
  \end{split}
\end{equation}
We carry out the $\int \d y$ integral analytically
\begin{equation}
  \label{eq:52}
  \int \d y \cdots = \frac{8}{3}\left[(2t-1)(t+1)^{1/2} -
    (2t+1)(t-1)^{1/2}\right],
\end{equation}
which asymptotes to $(4/3)t^{-3/2}$ when $t\gg 1$ (i.e., the wave amplitude is either very small $\xi\rightarrow0$ or very large $\xi \rightarrow\infty$) and hence $I(\xi\rightarrow 0) = 2/3$.

\begin{figure}
  \centering
\includegraphics[width = 0.45 \textwidth]{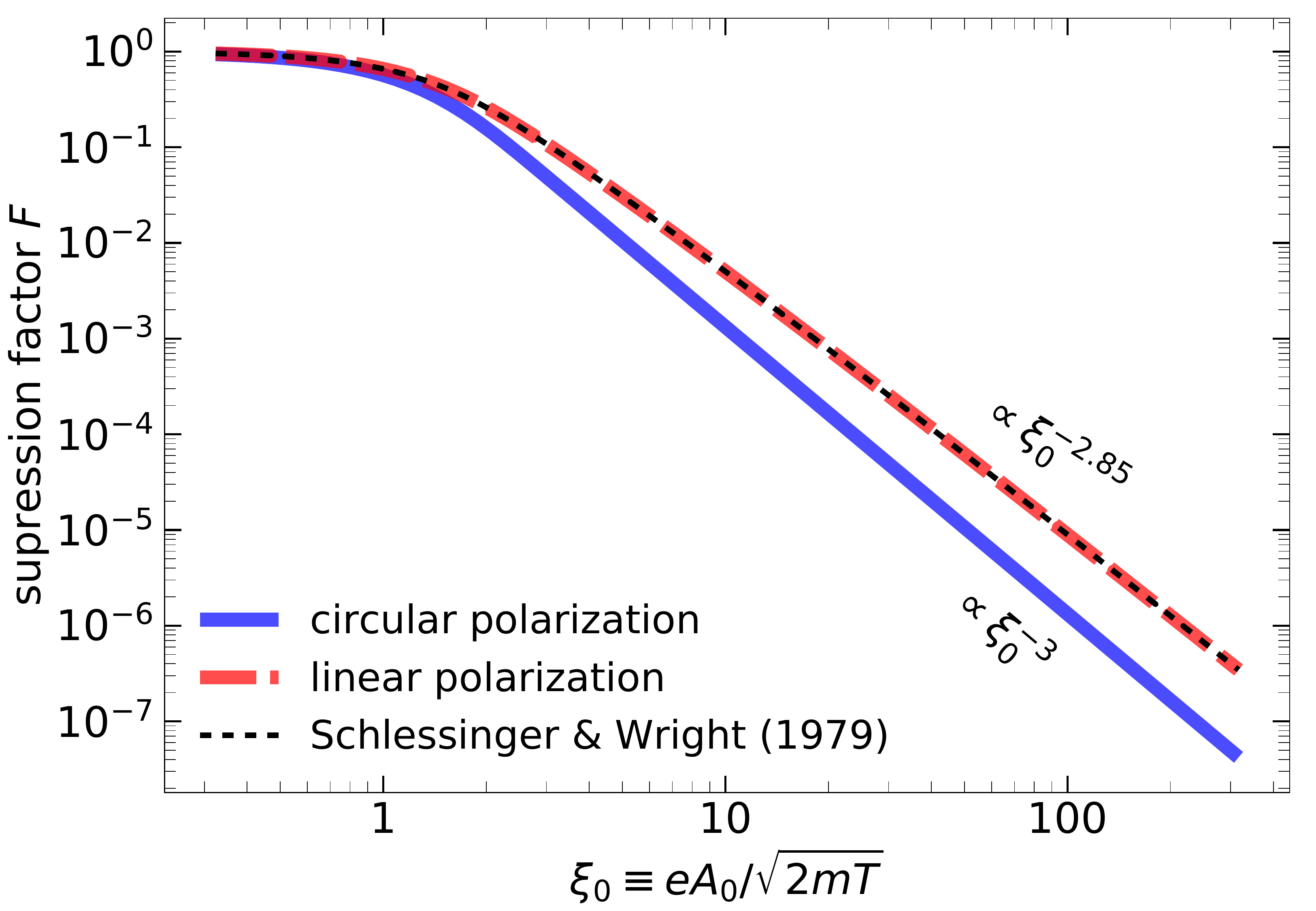}
\caption{The suppression factor $F(\xi_0)$ (eq. \ref{eq:53}) for ff absorption due to finite-amplitude waves, for circular polarization (blue solid line) and linear polarization (red dashed line), not including the Coulomb logarithm. The dimensionless  amplitude $\xi_0$ is roughly the ratio between the oscillatory speed and the thermal speed. In the large amplitude limit $\xi_0\gg 1$, the scalings are $F\propto \xi_0^{-3}$ for circular polarization and $\propto \xi_0^{-2.85}$ for linear polarization. Our linear polarization case agrees with \citet[][their eq. 3.34]{1979PhRvA..20.1934S} as shown in black dotted line.
}\label{fig:ffabs}
\end{figure}

The suppression factor at dimensionless wave amplitude $\xi_0$, as defined in equation (\ref{eq:41}), is given by
\begin{equation}
  \label{eq:53}
  F(\xi_0) \equiv \frac{3}{2\pi \lara{A^2}} {\int_0^\pi \d \psi\,
    A^2(\psi) I(\xi)},\
  \xi_0\equiv \frac{eA_0}{\sqrt{2mT}}.
\end{equation}
For circular polarization, the amplitude $A(\psi)=A_0$ is independent of phase, and for linear polarization, the amplitude is given by $A(\psi)=A_0\sin\psi$ (and hence $\lara{A^2} = A_0^2/2$). We show the numerical results of $F(\xi_0)$ in Fig. \ref{fig:ffabs}. In the limit $\xi_0\gg 1$, the suppression factor $F(\xi_0)$ scales as $\xi_0^{-3}$ for circular polarization and $\xi_0^{-2.85}$ for linear polarization. The slightly shallower behavior for linear polarization is because the oscillatory speed is much less than the maximum $eA_0/m$ for a fraction of the wave period, which leads to less suppression overall. We find that $F(\xi_0)$ is well approximated by equation (\ref{eq:55}).

\section{Summary}
The effects of finite-amplitude electromagnetic waves leave interesting imprints of the local ($\lesssim 1\rm\,pc$) environment of FRB sources on the DM, RM, and the ff absorption optical depth. We show that bursts with an isotropic luminosity of $10^{45}\rm\,erg\,s^{-1}$ will have DM modulation up to $10^{-2}\rm\,pc\,cm^{-3}$. We provide a classical calculation of how the ff absorption is affected by finite-amplitude effects. We find that sufficiently bright, low-frequency bursts are able to avoid ff absorption by an otherwise optically thick plasma. The DM/RM modulation and characteristic ff absorption frequency provide information on the density and magnetization of the plasma surrounding the FRB, and its distance from the source.

\section{acknowledgments}
We thank the anonymous referee whose suggestions have improved the clarity of this paper.
We acknowledge useful discussions with Peter Goldreich, Barak Zackay, Paz Beniamini, and Bing Zhang.

{\small
\bibliographystyle{mnras}
\bibliography{refs}

\begin{thebibliography}{}
\makeatletter
\relax
\def\mn@urlcharsother{\let\do\@makeother \do\$\do\&\do\#\do\^\do\_\do\%\do\~}
\def\mn@doi{\begingroup\mn@urlcharsother \@ifnextchar [ {\mn@doi@}
  {\mn@doi@[]}}
\def\mn@doi@[#1]#2{\def\@tempa{#1}\ifx\@tempa\@empty \href
  {http://dx.doi.org/#2} {doi:#2}\else \href {http://dx.doi.org/#2} {#1}\fi
  \endgroup}
\def\mn@eprint#1#2{\mn@eprint@#1:#2::\@nil}
\def\mn@eprint@arXiv#1{\href {http://arxiv.org/abs/#1} {{\tt arXiv:#1}}}
\def\mn@eprint@dblp#1{\href {http://dblp.uni-trier.de/rec/bibtex/#1.xml}
  {dblp:#1}}
\def\mn@eprint@#1:#2:#3:#4\@nil{\def\@tempa {#1}\def\@tempb {#2}\def\@tempc
  {#3}\ifx \@tempc \@empty \let \@tempc \@tempb \let \@tempb \@tempa \fi \ifx
  \@tempb \@empty \def\@tempb {arXiv}\fi \@ifundefined
  {mn@eprint@\@tempb}{\@tempb:\@tempc}{\expandafter \expandafter \csname
  mn@eprint@\@tempb\endcsname \expandafter{\@tempc}}}

\bibitem[\protect\citeauthoryear{{Akhiezer} \& {Polovin}}{{Akhiezer} \&
  {Polovin}}{1956}]{Akhiezer56}
{Akhiezer} A.,  {Polovin} R.,  1956, Soviet Phys. JETP, 3

\bibitem[\protect\citeauthoryear{{Bhandari} et~al.,}{{Bhandari}
  et~al.}{2018}]{2018MNRAS.475.1427B}
{Bhandari} S.,  et~al., 2018, \mn@doi [\mnras] {10.1093/mnras/stx3074}, \href
  {https://ui.adsabs.harvard.edu/abs/2018MNRAS.475.1427B} {475, 1427}

\bibitem[\protect\citeauthoryear{{Bornath}, {Schlanges}, {Hilse}  \&
  {Kremp}}{{Bornath} et~al.}{2001}]{2001PhRvE..64b6414B}
{Bornath} T.,  {Schlanges} M.,  {Hilse} P.,   {Kremp} D.,  2001, \mn@doi
  [Physical Review E] {10.1103/PhysRevE.64.026414}, \href
  {https://ui.adsabs.harvard.edu/abs/2001PhRvE..64b6414B} {64, 026414}

\bibitem[\protect\citeauthoryear{{CHIME/FRB Collaboration} et~al.,}{{CHIME/FRB
  Collaboration} et~al.}{2019}]{2019Natur.566..230C}
{CHIME/FRB Collaboration} et~al., 2019, \mn@doi [\nat]
  {10.1038/s41586-018-0867-7}, \href
  {https://ui.adsabs.harvard.edu/abs/2019Natur.566..230C} {566, 230}

\bibitem[\protect\citeauthoryear{{Chian}}{{Chian}}{1981}]{1981PhRvA..24.2773C}
{Chian} A.~C.~L.,  1981, \mn@doi [\pra] {10.1103/PhysRevA.24.2773}, \href
  {https://ui.adsabs.harvard.edu/abs/1981PhRvA..24.2773C} {24, 2773}

\bibitem[\protect\citeauthoryear{{Decker}, {Mori}, {Dawson}  \&
  {Katsouleas}}{{Decker} et~al.}{1994}]{1994PhPl....1.4043D}
{Decker} C.~D.,  {Mori} W.~B.,  {Dawson} J.~M.,   {Katsouleas} T.,  1994,
  \mn@doi [Physics of Plasmas] {10.1063/1.870874}, \href
  {https://ui.adsabs.harvard.edu/abs/1994PhPl....1.4043D} {1, 4043}

\bibitem[\protect\citeauthoryear{{Dobrowolny} \& {Ferrari}}{{Dobrowolny} \&
  {Ferrari}}{1976}]{1976A&A....47...97D}
{Dobrowolny} M.,  {Ferrari} A.,  1976, \aap, \href
  {https://ui.adsabs.harvard.edu/abs/1976A&A....47...97D} {47, 97}

\bibitem[\protect\citeauthoryear{{Farah} et~al.,}{{Farah}
  et~al.}{2019}]{2019MNRAS.488.2989F}
{Farah} W.,  et~al., 2019, \mn@doi [\mnras] {10.1093/mnras/stz1748}, \href
  {https://ui.adsabs.harvard.edu/abs/2019MNRAS.488.2989F} {488, 2989}

\bibitem[\protect\citeauthoryear{{Ferrari}, {Massaglia}  \&
  {Dobrowolny}}{{Ferrari} et~al.}{1975}]{1975PhLA...55..227F}
{Ferrari} A.,  {Massaglia} S.,   {Dobrowolny} M.,  1975, \mn@doi [Physics
  Letters A] {10.1016/0375-9601(75)90722-7}, \href
  {https://ui.adsabs.harvard.edu/abs/1975PhLA...55..227F} {55, 227}

\bibitem[\protect\citeauthoryear{{Gourdji}, {Michilli}, {Spitler}, {Hessels},
  {Seymour}, {Cordes}  \& {Chatterjee}}{{Gourdji}
  et~al.}{2019}]{2019ApJ...877L..19G}
{Gourdji} K.,  {Michilli} D.,  {Spitler} L.~G.,  {Hessels} J.~W.~T.,  {Seymour}
  A.,  {Cordes} J.~M.,   {Chatterjee} S.,  2019, \mn@doi [\apjl]
  {10.3847/2041-8213/ab1f8a}, \href
  {https://ui.adsabs.harvard.edu/abs/2019ApJ...877L..19G} {877, L19}

\bibitem[\protect\citeauthoryear{{Gunn} \& {Ostriker}}{{Gunn} \&
  {Ostriker}}{1971}]{1971ApJ...165..523G}
{Gunn} J.~E.,  {Ostriker} J.~P.,  1971, \mn@doi [\apj] {10.1086/150919}, \href
  {https://ui.adsabs.harvard.edu/abs/1971ApJ...165..523G} {165, 523}

\bibitem[\protect\citeauthoryear{{Kaw} \& {Dawson}}{{Kaw} \&
  {Dawson}}{1970}]{1970PhFl...13..472K}
{Kaw} P.,  {Dawson} J.,  1970, \mn@doi [Physics of Fluids] {10.1063/1.1692942},
  \href {https://ui.adsabs.harvard.edu/abs/1970PhFl...13..472K} {13, 472}

\bibitem[\protect\citeauthoryear{{Kroll} \& {Watson}}{{Kroll} \&
  {Watson}}{1973}]{1973PhRvA...8..804K}
{Kroll} N.~M.,  {Watson} K.~M.,  1973, \mn@doi [Physical Review A]
  {10.1103/PhysRevA.8.804}, \href
  {https://ui.adsabs.harvard.edu/abs/1973PhRvA...8..804K} {8, 804}

\bibitem[\protect\citeauthoryear{{Law} et~al.,}{{Law}
  et~al.}{2017}]{2017ApJ...850...76L}
{Law} C.~J.,  et~al., 2017, \mn@doi [\apj] {10.3847/1538-4357/aa9700}, \href
  {https://ui.adsabs.harvard.edu/abs/2017ApJ...850...76L} {850, 76}

\bibitem[\protect\citeauthoryear{{Lorimer}, {Bailes}, {McLaughlin}, {Narkevic}
  \& {Crawford}}{{Lorimer} et~al.}{2007}]{2007Sci...318..777L}
{Lorimer} D.~R.,  {Bailes} M.,  {McLaughlin} M.~A.,  {Narkevic} D.~J.,
  {Crawford} F.,  2007, \mn@doi [Science] {10.1126/science.1147532}, \href
  {https://ui.adsabs.harvard.edu/abs/2007Sci...318..777L} {318, 777}

\bibitem[\protect\citeauthoryear{{Lu} \& {Piro}}{{Lu} \&
  {Piro}}{2019}]{2019ApJ...883...40L}
{Lu} W.,  {Piro} A.~L.,  2019, \mn@doi [\apj] {10.3847/1538-4357/ab3796}, \href
  {https://ui.adsabs.harvard.edu/abs/2019ApJ...883...40L} {883, 40}

\bibitem[\protect\citeauthoryear{{Luan} \& {Goldreich}}{{Luan} \&
  {Goldreich}}{2014}]{2014ApJ...785L..26L}
{Luan} J.,  {Goldreich} P.,  2014, \mn@doi [\apjl]
  {10.1088/2041-8205/785/2/L26}, \href
  {https://ui.adsabs.harvard.edu/abs/2014ApJ...785L..26L} {785, L26}

\bibitem[\protect\citeauthoryear{{Max} \& {Perkins}}{{Max} \&
  {Perkins}}{1971}]{1971PhRvL..27.1342M}
{Max} C.,  {Perkins} F.,  1971, \mn@doi [\prl] {10.1103/PhysRevLett.27.1342},
  \href {https://ui.adsabs.harvard.edu/abs/1971PhRvL..27.1342M} {27, 1342}

\bibitem[\protect\citeauthoryear{{Max} \& {Perkins}}{{Max} \&
  {Perkins}}{1972}]{1972PhRvL..29.1731M}
{Max} C.,  {Perkins} F.,  1972, \mn@doi [\prl] {10.1103/PhysRevLett.29.1731},
  \href {https://ui.adsabs.harvard.edu/abs/1972PhRvL..29.1731M} {29, 1731}

\bibitem[\protect\citeauthoryear{{Mourou}, {Tajima}  \& {Bulanov}}{{Mourou}
  et~al.}{2006}]{2006RvMP...78..309M}
{Mourou} G.~A.,  {Tajima} T.,   {Bulanov} S.~V.,  2006, \mn@doi [Reviews of
  Modern Physics] {10.1103/RevModPhys.78.309}, \href
  {https://ui.adsabs.harvard.edu/abs/2006RvMP...78..309M} {78, 309}

\bibitem[\protect\citeauthoryear{{Noerdlinger}}{{Noerdlinger}}{1971}]{1971PhFl...14..999N}
{Noerdlinger} P.,  1971, \mn@doi [Physics of Fluids] {10.1063/1.1693561}, \href
  {https://ui.adsabs.harvard.edu/abs/1971PhFl...14..999N} {14, 999}

\bibitem[\protect\citeauthoryear{{Pert}}{{Pert}}{1995}]{1995PhRvE..51.4778P}
{Pert} G.~J.,  1995, \mn@doi [\pre] {10.1103/PhysRevE.51.4778}, \href
  {https://ui.adsabs.harvard.edu/abs/1995PhRvE..51.4778P} {51, 4778}

\bibitem[\protect\citeauthoryear{{Prochaska} et~al.,}{{Prochaska}
  et~al.}{2019}]{2019Sci...366..231P}
{Prochaska} J.~X.,  et~al., 2019, \mn@doi [Science] {10.1126/science.aay0073},
  \href {https://ui.adsabs.harvard.edu/abs/2019Sci...366..231P} {366, 231}

\bibitem[\protect\citeauthoryear{{Ravi}}{{Ravi}}{2019}]{2019MNRAS.482.1966R}
{Ravi} V.,  2019, \mn@doi [\mnras] {10.1093/mnras/sty1551}, \href
  {https://ui.adsabs.harvard.edu/abs/2019MNRAS.482.1966R} {482, 1966}

\bibitem[\protect\citeauthoryear{{Ravi} \& {Loeb}}{{Ravi} \&
  {Loeb}}{2019}]{2019ApJ...874...72R}
{Ravi} V.,  {Loeb} A.,  2019, \mn@doi [\apj] {10.3847/1538-4357/ab0748}, \href
  {https://ui.adsabs.harvard.edu/abs/2019ApJ...874...72R} {874, 72}

\bibitem[\protect\citeauthoryear{{Ravi} et~al.,}{{Ravi}
  et~al.}{2019}]{2019Natur.572..352R}
{Ravi} V.,  et~al., 2019, \mn@doi [\nat] {10.1038/s41586-019-1389-7}, \href
  {https://ui.adsabs.harvard.edu/abs/2019Natur.572..352R} {572, 352}

\bibitem[\protect\citeauthoryear{{Rybicki} \& {Lightman}}{{Rybicki} \&
  {Lightman}}{1979}]{1979rpa..book.....R}
{Rybicki} G.~B.,  {Lightman} A.~P.,  1979, {Radiative processes in
  astrophysics}.
John Wiley and Sons

\bibitem[\protect\citeauthoryear{{Schlessinger} \& {Wright}}{{Schlessinger} \&
  {Wright}}{1979}]{1979PhRvA..20.1934S}
{Schlessinger} L.,  {Wright} J.,  1979, \mn@doi [Physical Review A]
  {10.1103/PhysRevA.20.1934}, \href
  {https://ui.adsabs.harvard.edu/abs/1979PhRvA..20.1934S} {20, 1934}

\bibitem[\protect\citeauthoryear{{Wang} \& {Zhang}}{{Wang} \&
  {Zhang}}{2019}]{2019ApJ...882..108W}
{Wang} F.~Y.,  {Zhang} G.~Q.,  2019, \mn@doi [\apj] {10.3847/1538-4357/ab35dc},
  \href {https://ui.adsabs.harvard.edu/abs/2019ApJ...882..108W} {882, 108}

\bibitem[\protect\citeauthoryear{{Yang} \& {Zhang}}{{Yang} \&
  {Zhang}}{2020}]{yangzhang20}
{Yang} Y.-P.,  {Zhang} B.,  2020, \mn@doi [\apjl] {10.3847/2041-8213/ab7ccf},
  \href {https://ui.adsabs.harvard.edu/abs/2020ApJ...892L..10Y} {892, L10}

\makeatother
\end{thebibliography}
}

\label{lastpage}
\end{document}